\documentclass[twoside]{article} % arXiv style

\usepackage[bindingoffset=0.in, left=1.in,right=1.in,top=0.75in,bottom=1in, footskip=.25in, headsep=.2in]{geometry} % margins
\usepackage{abstract}
\usepackage{authblk}
\usepackage{blindtext}
\usepackage{fancyhdr}
\usepackage{sectsty}

\sectionfont{\fontsize{12}{15}\selectfont}
\subsectionfont{\fontsize{10}{15}\selectfont}

\usepackage[american]{babel}% hyphenation rules
\addto{\captionsamerican}{}

\usepackage{graphicx}
\usepackage{amsmath}
\usepackage{amssymb} % Mathematical set symbols
\usepackage{upgreek}
\usepackage[normalem]{ulem}
\usepackage{bm} % Bold math
\usepackage[utf8]{inputenc} % Allow for accents
\usepackage[american]{babel} % hyphenation rules
\usepackage[bookmarks=false]{hyperref}
\hypersetup{colorlinks={true},linkcolor={blue},citecolor={blue},urlcolor={blue}}
\usepackage{lscape}

\usepackage{siunitx}

\usepackage{textcomp,gensymb}

\usepackage[square]{natbib}

\usepackage{enumitem}

\usepackage{physics}

\newcommand{\pd}[2]{\partial_{#2}{#1}} % partial derivatives
\newcommand{\td}[2]{\frac{\textrm{d}#1}{\textrm{d}#2}} % total derivatives
\newcommand{\md}[2]{\frac{\textrm{D}#1}{\textrm{D}#2}} % material derivatives long
\newcommand{\unitvect}[1]{\widehat{\mathbf{#1}}}
\newcommand{\vect}[1]{\mathbf{#1}}
\newcommand{\boldnabla}{\boldsymbol{\nabla}} % bold nabla
\def \reff#1 {
	(\ref{#1})
}

\usepackage{float}
\usepackage{placeins}
\usepackage{caption}

\usepackage{lineno} % new -- allow for line numbering
\def \fig [#1][#2]#3#4 {
  \begin{figure}
    \centering
      \includegraphics[width=#2\textwidth]{figs/#1}
      \caption{#3}{\footnotesize{#4}}
      \label{fig:#1}
  \end{figure}
}

\def \subfigs [#1]#2#3#4 {
  \begin{figure}
    \centering
      #2
      \caption{#3}{\footnotesize{#4}}
      \label{fig:#1}
  \end{figure}
}

\def \subfig [#1][#2]#3 {
  \begin{subfigure}[b]{#2\textwidth}
    \includegraphics[width=\textwidth]{figs/#1}
    \caption{#3}
    \label{fig:#1}
  \end{subfigure}
}

\usepackage{etoolbox}

\makeatletter

\patchcmd{\NAT@citex}
  {\@citea\NAT@hyper@{%
     \NAT@nmfmt{\NAT@nm}%
     \hyper@natlinkbreak{\NAT@aysep\NAT@spacechar}{\@citeb\@extra@b@citeb}%
     \NAT@date}}
  {\@citea\NAT@nmfmt{\NAT@nm}%
   \NAT@aysep\NAT@spacechar\NAT@hyper@{\NAT@date}}{}{}

\patchcmd{\NAT@citex}
  {\@citea\NAT@hyper@{%
     \NAT@nmfmt{\NAT@nm}%
     \hyper@natlinkbreak{\NAT@spacechar\NAT@@open\if*#1*\else#1\NAT@spacechar\fi}%
       {\@citeb\@extra@b@citeb}%
     \NAT@date}}
  {\@citea\NAT@nmfmt{\NAT@nm}%
   \NAT@spacechar\NAT@@open\if*#1*\else#1\NAT@spacechar\fi\NAT@hyper@{\NAT@date}}
  {}{}
  
\makeatother

\usepackage{booktabs}

\title{\bf Seagrass deformation affects fluid instability\\ and tracer exchange in canopy flow}

\author{Guilherme S. Vieira\,$^{1,}$\thanks{salvadorvieira.g@northeastern.edu} }
\author{Michael R. Allshouse\,$^{1,}$\thanks{m.allshouse@northeastern.edu} }
\author{Amala Mahadevan\,$^{2,}$\thanks{amala@whoi.edu} }
\affil{$^{1}$\,\normalsize{Department of Mechanical and Industrial Engineering, Northeastern University,\\}
\normalsize{Boston, MA 02115, USA\\}
$^{2}$\,\normalsize{Department of Physical Oceanography, Woods Hole Oceanographic Institution,\\}
\normalsize{Woods Hole, MA 02543, USA}\\
\vspace{1em}
Submitted to {\it PNAS} on January 20, 2022
}

\date{}

\begin{document}

\renewcommand{\abstractname}{}    % clear the title
\renewcommand{\absnamepos}{empty} % originally center
 
\maketitle

\vspace{-2em}

% \keywords{Keyword1, Keyword2, Keyword3}

%%%%%%%%%%%%%%%%%%%%%%%%%%%%%%%%%%%%%%%%%%
% Abstract (Do not insert blank lines, i.e. \\) 
\begin{abstract} % (207 words)

\begin{center}
    \textbf{Abstract}
\end{center}
\vspace{0.5em}

Monami is the synchronous waving of a submerged seagrass bed in response to unidirectional fluid flow. Here we develop a multiphase model for the dynamical instabilities and flow-driven collective motions of buoyant, deformable seagrass. We show that the impedance to flow due to the seagrass results in an unstable velocity shear layer at the canopy interface, leading to a periodic array of vortices that propagate downstream. Each passing vortex locally weakens the along-stream velocity at the canopy top, reducing the drag and allowing the deformed grass to straighten up just beneath it. This causes the grass to oscillate periodically. Crucially, the maximal grass deflection is out of phase with the vortices. A phase diagram for the onset of instability shows its dependence on the fluid Reynolds number and an effective buoyancy parameter. Less buoyant grass is more easily deformed by the flow and forms a weaker shear layer, with smaller vortices and less material exchange across the canopy top. While higher Reynolds number leads to stronger vortices and larger waving amplitudes of the seagrass, waving is maximized at intermediate grass buoyancy. All together, our theory and computations correct some misconceptions in interpretation of the mechanism and provide a robust explanation consistent with a number of experimental observations.

\vspace{1em}
\begin{center}
    \textbf{Significance Statement}
\end{center}
\vspace{0.25em}

Seagrass meadows serve as breeding grounds for marine organisms and as blue carbon repositories. Flow through submerged seagrass can lead to the synchronous waving of the grass, a phenomenon known as monami that has been explored in a number of experimental studies. Limitations in visualizing the entire flow field as it interacts with the grass blades, however, leave aspects of the phenomenon in need of better explanation. By developing a coupled fluid-structure model for monami, we perform numerical simulations of the fluid dynamical instability, vortex formation, and seagrass waving for a range of parameters. We explore the dependence of instability, flow structures, grass deformation and material exchange on the Reynolds number and grass buoyancy.

\end{abstract}

%%%%%%%%%%%%%%%%%%%%%%%%%%%%%%%%%%%%%%%%%%%%%%%%%%%%%%%%%%
% Introduction
\section{Introduction}
\label{sec:Intro}

Seagrass is typically deformable, which allows the grass blades to reconfigure according to the fluid load~\citep{vogel2020life}.
While emergent canopies -- those that are in the inter-tidal zone and emerge above the water surface -- need stiffness for the stems to stand up out of the water, fully submerged seagrass species (such as \emph{Halodule wrightii} and \emph{Syringodium filiforme}) tend to stand up by buoyancy~\citep{wilson2010seagrass}.
In order to photosynthesize, submerged canopies have a typical height comparable to the water depth~\citep{marion2014aquatic}, which results in a significant portion of the flow being obstructed by the canopy. Seagrass beds exhibit a particularly rich set of dynamic behaviors due to their collective interaction with the flow.
Hydrodynamic processes resulting from these interactions influence environmental processes such as sedimentation, transport of dissolved oxygen~\citep{long2020ebullition} and nutrients, plant growth, and biomass production~\citep{fonseca1987effects,grizzle1996hydrodynamically,nepf1999drag,nepf2012flow}.
Seagrass meadows are also believed to influence sediment deposition and resuspension~\citep{short1984seagrass,walker1996experimental}, as vegetation can trap suspended materials~\citep{short1984seagrass} and reduce sediment movement~\citep{fonseca1986comparison}.

Instabilities of flow through submerged canopies yield a phenomenon known as \emph{monami} -- the progressive, synchronous oscillation of aquatic vegetation~\citep{ackerman1993reduced,nepf2012flow}.
Current explanations of monami~\citep{ikeda1996three,raupach1996coherent,ghisalberti2002mixing} rely on the existence of a shear layer at the top of the grass bed due to vegetation drag.
Through a mechanism similar to the Kelvin–Helmholtz instability~\citep{singh2016linear}, the enhanced velocity shear near the grass top creates a sheet of vorticity that destabilizes into vortices over time.
These vortices perturb the flow, which locally changes the deformation of grass blades and leads to synchronous oscillations of the grass bed. 
These perturbations to the mean flow have been observed experimentally and feature sweeps and ejections that occur at the leading and trailing edges of vortices, respectively~\citep{ghisalberti2006structure,nezu2008turbulence,okamoto2009turbulence}.
Transport of material across the canopy has also been studied experimentally ~\citep{ghisalberti2005mass, nepf2008flow}. Our numerical simulations provide a complementary and comprehensive picture of the fluid instability, vortex-seagrass interaction and tracer exchange between the seagrass bed and the overflow, in terms of its dependence on seagrass buoyancy and Reynolds number.

There are numerous modeling challenges in capturing the properties of this system, primarily related to the feedback mechanism between flow and vegetation.
In a two-way coupled dynamic model, the fluid will apply a load on each vegetative structure, which causes a resultant deformation that, in turn, affects the flow~\citep{de2008effects}.
Thus, in general, the fluid flow must be solved simultaneously with the configuration of each structure.
These challenges have demanded sophisticated studies, both experimental~\citep{dunn1996mean,ghisalberti2004limited,okamoto2009turbulence,hu2014laboratory,mandel2019surface} and numerical~\citep{dupont2010modelling,zeller2015simple,beudin2017development,mattis2019computational,sundin2019interaction}.
Most previous simplified models fall into one of two categories: models of flow over a specified set of rigid obstacles~\citep{ghisalberti2004limited,singh2016linear}, or models where grass deformation can occur, but does not change the flow profile~\citep{luhar2011flow}.
Fewer models~\citep{wong2020shear} emphasize the coupling between grass deformation and flow, and our study is unique as it presents numerical simulations of the coupled system and uses them to study monami.

\begin{figure}[t!]
    \centering
    \includegraphics[width=0.975\textwidth]{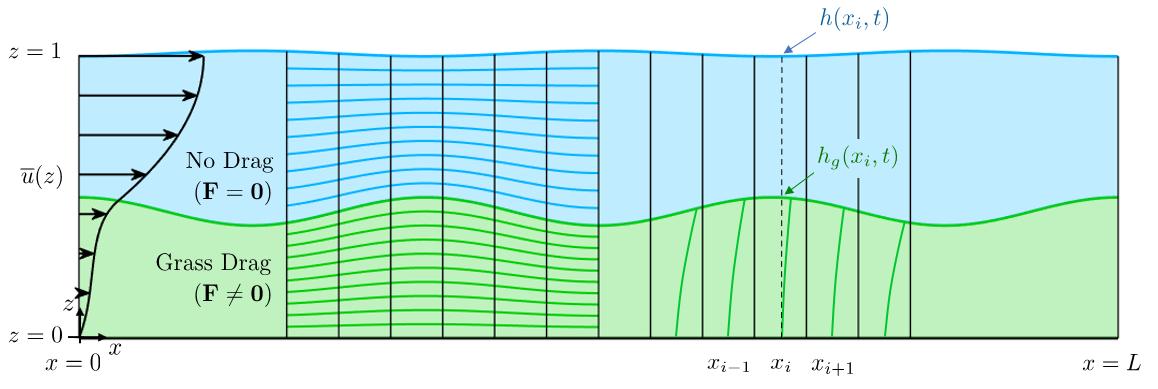}
    \caption{Schematic of the domain used in the simulations.
    The steady-state horizontal velocity profile $\overline{u}(z)$ is imposed as the velocity inlet boundary condition. A conformal map accounts for variations in $h$ and $h_g$ to separate the overflow (where $\mathbf{F}=\mathbf{0}$, in blue) and seagrass (where $\mathbf{F}\neq\mathbf{0}$, in green) regions of the domain. Buoyant grass blades deform by the flow, apply a drag on the fluid $\mathbf{F}$, and the composite tip positions determine $h_g$.
    }
    \label{fig:full_domain_intro}
\end{figure}

This study builds on previous work by \citet{singh2016linear} and \citet{wong2020shear} in analyzing the dynamics of flow through a submerged seagrass canopy and its resultant instabilities.
Although monami is manifested in the grass motion, the drag exerted by the vegetation on the flow is central to the instability, and the resulting flow structures persist in laboratory experiments even when deformable grass is replaced by rigid dowels~\citep{ghisalberti2002mixing,ghisalberti2006structure}.
\citet{singh2016linear} proposed the seagrass effect on the fluid to be modeled as a continuum drag acting perpendicular to the blade, proportional to the number of stems per unit area, and established the dependence between viscous effects and flow instabilities by performing a linear stability analysis of flows through an array of rigid beams. 
\citet{wong2020shear} expanded this model to account for flexible beams, derived the coupled equations of motion and relevant dimensionless groups, and performed a stability analysis to investigate conditions for the onset of instabilities.

We are interested in the impact that the grass blade deflection has on the onset of the instability, progression of the developed vortices, and material transport resulting from this interaction. 
Our model incorporates blade deformability into the two-phase model by~\citet{singh2016linear}, but as opposed to the approach adopted by~\citet{wong2020shear}, where the grass blades are modeled as linearly elastic flexible beams with one end clamped perpendicularly to the seabed, in our model the submerged grass blades stand up by buoyancy, do not resist shear (zero flexural rigidity), and are always in equilibrium with the flow (no contribution of the inertial term in the equations of motion for the grass). 
These assumptions simplify the equations of motion for the grass, while successfully reproducing the monami dynamics.

To model a submerged seagrass bed, we solve the Navier-Stokes equations for two phases: the grass-free overflow, and the grass-bed in which the seagrass contributes a bulk volumetric drag $\mathbf{F}$ that depends on the blade positions and velocity field (Fig.~\ref{fig:full_domain_intro}). 
The drag is quadratic in the velocity normal to the grass blades and hence depends on the grass shape, which in turn depends on the fluid drag. 
We model the shape of representative grass blades rooted to the bed in the center of each grid cell column (in plan view) by assuming a balance between drag, which deforms the grass blade, and buoyancy, which restores its shape to vertical.  
There are $\overline{N}$ grass blades per unit area that impose drag on the fluid, but do not block the flow.

Simulations are performed using a version of the non-hydrostatic Process Study Ocean Model (PSOM)~\citep{mahadevan1996a_nonhydrostatic,mahadevan1996b_nonhydrostatic}. 
The submerged seagrass bed of undisturbed height $\ell$ is modeled in an open channel of undisturbed water height $H$ using a grid that conforms to the free surface $h(x,t)$ and seagrass height $h_g(x,t)$ as seen in Fig.~\ref{fig:full_domain_intro}. 
The along-channel coordinate is $x$, the vertical coordinate is $z$, and for the study described here, variations in cross-channel ($y$) direction are set to zero. 
The inflow velocity profile is in equilibrium with the grass, and within a buffer of the outflow boundary, we restore the velocity profile to the same equilibrium profile. 
All variables are non-dimensionalized using the undisturbed water height $H$ as the characteristic length scale, the horizontal flow speed at the free surface $U$ as the velocity scale, and $H/U$ as the timescale. 
Variables are henceforth presented in dimensionless form. 
The dimensionless parameters that govern the solution are
\begin{linenomath*}
\begin{equation}
Re=UH/\nu^*, \quad \beta=\frac{(\rho-\rho_g)gd}{\rho c_D U^2}, \quad
r = \ell/H, \quad
\lambda=c_D\overline{N}bH, \quad
Fr=U/\sqrt{gH}.
\label{eq:nd-parameters}
\end{equation}
\end{linenomath*}
These are similar to \citet{wong2020shear}, except for $\beta$.
Here,  $\nu^*$ is the constant eddy viscosity, $\rho$ is the fluid density, $\rho_g$ is the grass density, $g$ is acceleration due to gravity, $c_D$ is the quadratic drag coefficient, $d$ is the thickness of the grass blades in the along-flow ($x$) direction, while $b$ is the width of the grass blades (in the $y$-direction). 
The Reynolds ($Re$) and Froude ($Fr$) numbers are standard parameters. The height ratio ($r$) is chosen as 0.5 in all our simulations, and this does not play a dominant role in any of the overall observations presented in this paper. The parameter $\lambda$ governs the drag impedance by the seagrass and affects the velocity shear and is chosen as 1. The buoyancy parameter $\beta$ is the ratio of seagrass buoyancy to drag and influences the shape of the grass blades. More buoyant grass has larger $\beta$ and deforms less due to the flow. Choosing $b$ and $d$ as independent parameters enables us to change $\beta$ without affecting $\lambda$. In this study, we perform numerical experiments for a range of $Re$, which is varied by changing $\nu^*$, and a range of $\beta$, which is varied by changing $d$.
We analyze the onset of flow instability and amplitude of grass oscillations as a function of Reynolds number and grass deformation. 
We examine how the vortex position aligns with the shape of the grass bed and how the exchange of material across the grass canopy is affected by these dynamics.

%%%%%%%%%%%%%%%%%%%%%%%%%%%%%%%%%%%%%%%%%%%%%%%%%%%%%%%%%%
% Results

\section{Results}
\label{sec:Results}

%%%% Subsection %%%%
%\subsection{Steady-state solution}
% \subsection{Prior to instability, }
% \label{subsec:single_blade}

%%% Subsection %%%
\subsection{Instability onset and progression}
\label{subsec:instability_onset}

Before the instability onset, the flow and grass are both steady.  The steady-state solution is a function of $z$ alone, and can be calculated with a simplified one-dimensional coupled model that eliminates dependence in $x$ and $t$ (Methods section). 

\begin{figure}[t!]
    \centering
    \includegraphics[width=0.65\textwidth]{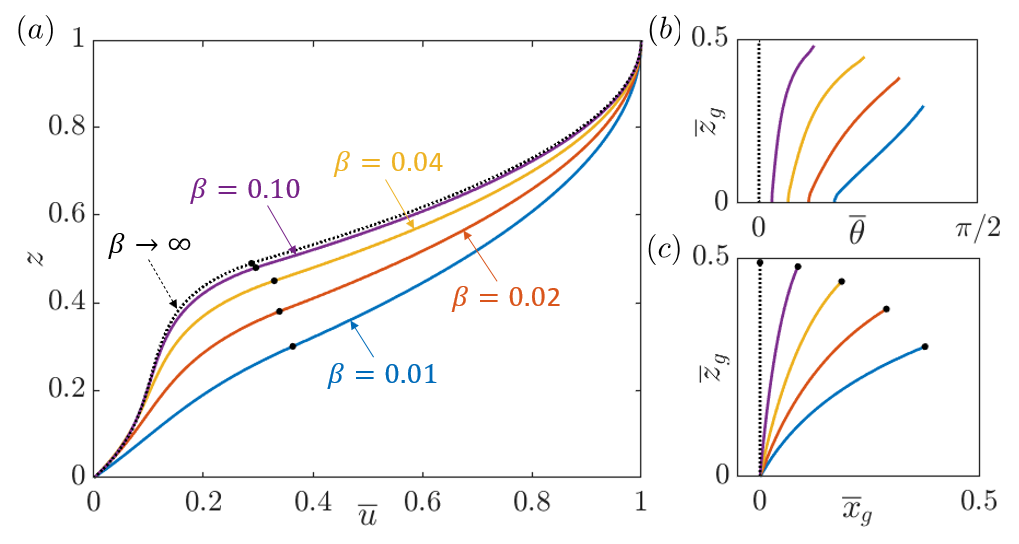}
    \caption{Steady-state $(a)$ horizontal velocity profile $\overline{u}(z)$, and the corresponding $(b)$ angle with the vertical $\overline{\theta}(z)$ and $(c)$ blade shape $(\overline{x}_g,\, \overline{z}_g)$, for $Re=1000$, $r=0.5$, $\lambda=1$, and $Fr^2=0.1$. Buoyant grass model, with no-slip velocity boundary condition at $z=0$, and $\beta=0.01$, 0.02, 0.04, 0.10, and $\infty$ (dotted). The black markers on $(a)$ and $(c)$ mark the vertical position of the corresponding canopy tip $\overline{h}_g$ in each case.}
    \label{fig:single_stem_noslip}
\end{figure}

The steady-state velocity profile $\overline{u}(z)$, grass shape $(\overline{x}_g,\,\overline{z}_g)$, and corresponding blade angle with the vertical $\overline{\theta}(z)$,  calculated for $Re=1000$, $r=0.5$, $\lambda= 1$, $Fr^2=0.1$,
and a range of values of the buoyancy parameter $\beta$, which in our model quantifies to what extent the blade can deform, is presented in Fig.~\ref{fig:single_stem_noslip}. 
Solutions are computed with fluid boundary conditions $\overline{u}=0$ (no-slip) at the bottom ($z=0$)  and $\dv*{\overline{u}}{z}=0$ at the surface  ($z=1$). The horizontal pressure gradient is adjusted so that $\overline{u}=1$ at the surface.  For the grass, the tension is zero at the tip, and the position is fixed at the bottom. The overbar is used to represent the steady-state solutions, which are independent of $x$ and $t$.

Whether the grass relies on bending stiffness (as in~\citet{wong2020shear}) or buoyancy (this study) to restore its shape, the shape of the deformed grass and its implication for flow instability and fluid exchange are qualitatively similar (Supplementary information). 
The smaller $\beta$, the more blade deflection, the larger the angle $\overline{\theta}$ along the blade, and the smaller the steady-state height $\overline{h}_g$ corresponding to the height of the tip. 
As $\beta$ increases, the velocity shear  $\dv*{\overline{u}}{z}$ at the canopy top ($z=\overline{h}_g$) monotonically grows, with the limiting case $\beta\to\infty$ corresponding to a fixed, vertical blade, and maximum shear. 

\begin{figure}[t!]
    \centering
    \includegraphics[width=\textwidth]{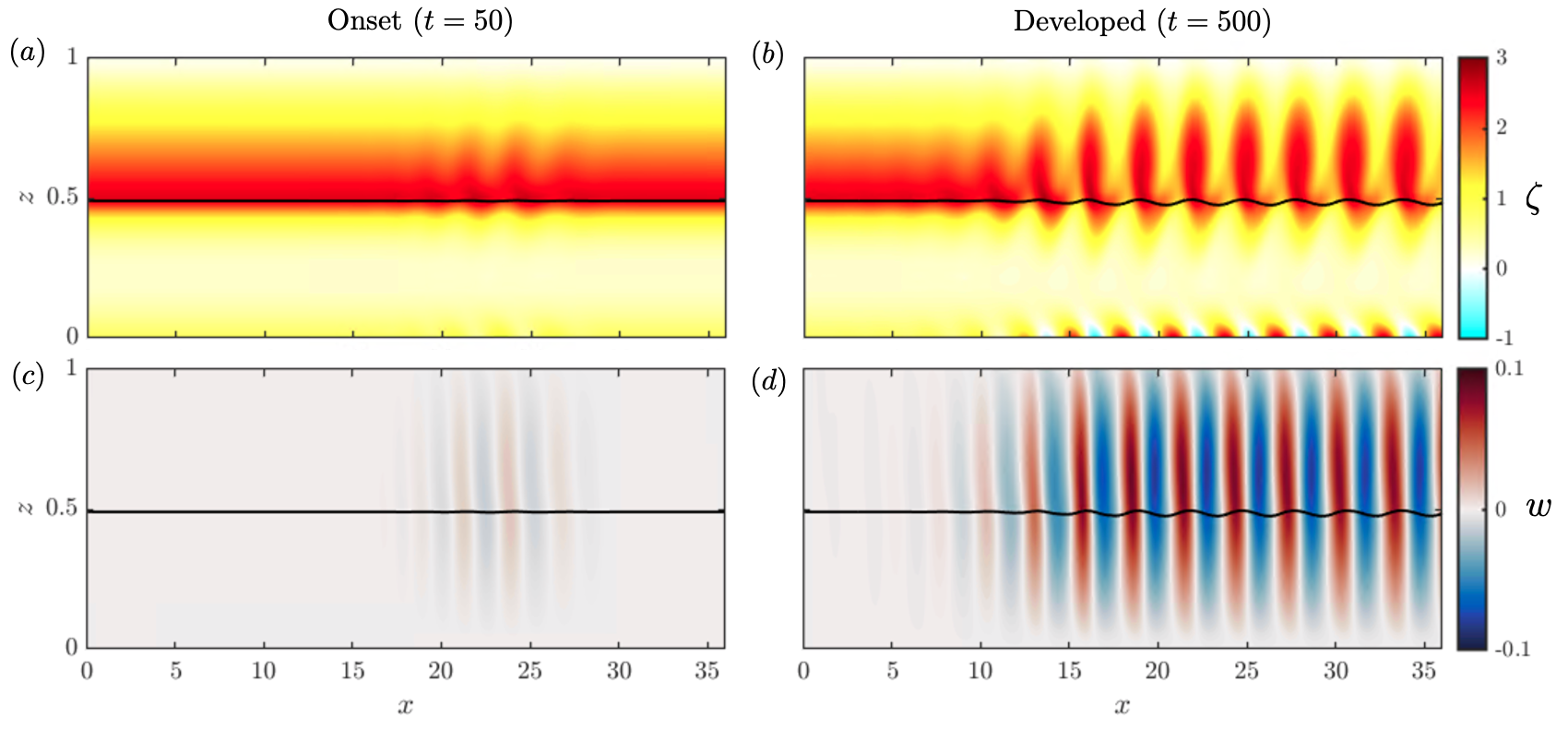}
    \caption{Instability onset at $t=50$ $(a, c)$, and developed, long-term behavior at $t=500$ $(b, d)$, for $Re=1000$, $\beta=0.10$, $r=0.5$, $\lambda=1$, and $Fr^2=0.1$. $(a,b)$ Vorticity field $\zeta=\pd{u}{z}-\pd{w}{x}$, and $(c,d)$ vertical velocity $w$. The solid black lines represent the instantaneous seagrass height $h_g(x,t)$.}
    \label{fig:result_onsetVisualization_field}
\end{figure}

We initialize the channel model with the steady-state solution described above, and with no vertical velocity. 
For sufficiently large $Re$ and $\lambda$, the shear layer at the canopy top is unstable, and instabilities are triggered spontaneously after finite time. 
The simulation (Fig.~\ref{fig:result_onsetVisualization_field}) with $Re=1000$, $r=0.5$, $\lambda=1$, $Fr^2=0.1$, and $\beta=0.10$ exhibits the instability onset at $t=50$ as seen in the vorticity $\zeta=\pd{u}{z}-\pd{w}{x}$  and vertical velocity $w$ fields (Figs~\ref{fig:result_onsetVisualization_field}$(a,c)$). In most of the domain, $w \approx 0$ and $\zeta \approx \dv*{\overline{u}}{z}$ with maximum $\zeta$ right above the canopy top, but some oscillations in $\zeta$ for $x\in[16,30]$, where shear-instabilities start to grow and induce alternating vertical velocities.

When the instability is fully developed at $t=500$ (Figs~\ref{fig:result_onsetVisualization_field}$(b,d)$), the vorticity rolls up to form vortices. Vortices are shed from $x\approx 7$,  grow until $x\approx 16$, and stabilize in size as they propagate downstream with the flow (time evolution video in Supplementary information).
The vortex centers lie between maxima and minima of vertical velocity, which peak just above the canopy (Fig.~\ref{fig:result_onsetVisualization_field}$(d)$).
We also observe alternating vorticity maxima and minima near the seabed at $z=0$ (Fig.~\ref{fig:result_onsetVisualization_field}($b$)), which indicates that flow perturbations induced by the vortices penetrate to the bottom and cause flow reversal ($u<0$) near the seabed, where the unperturbed velocity is already small due to the no-slip bottom boundary condition. 

\begin{figure}[t!]
    \centering
    \includegraphics[width=\textwidth]{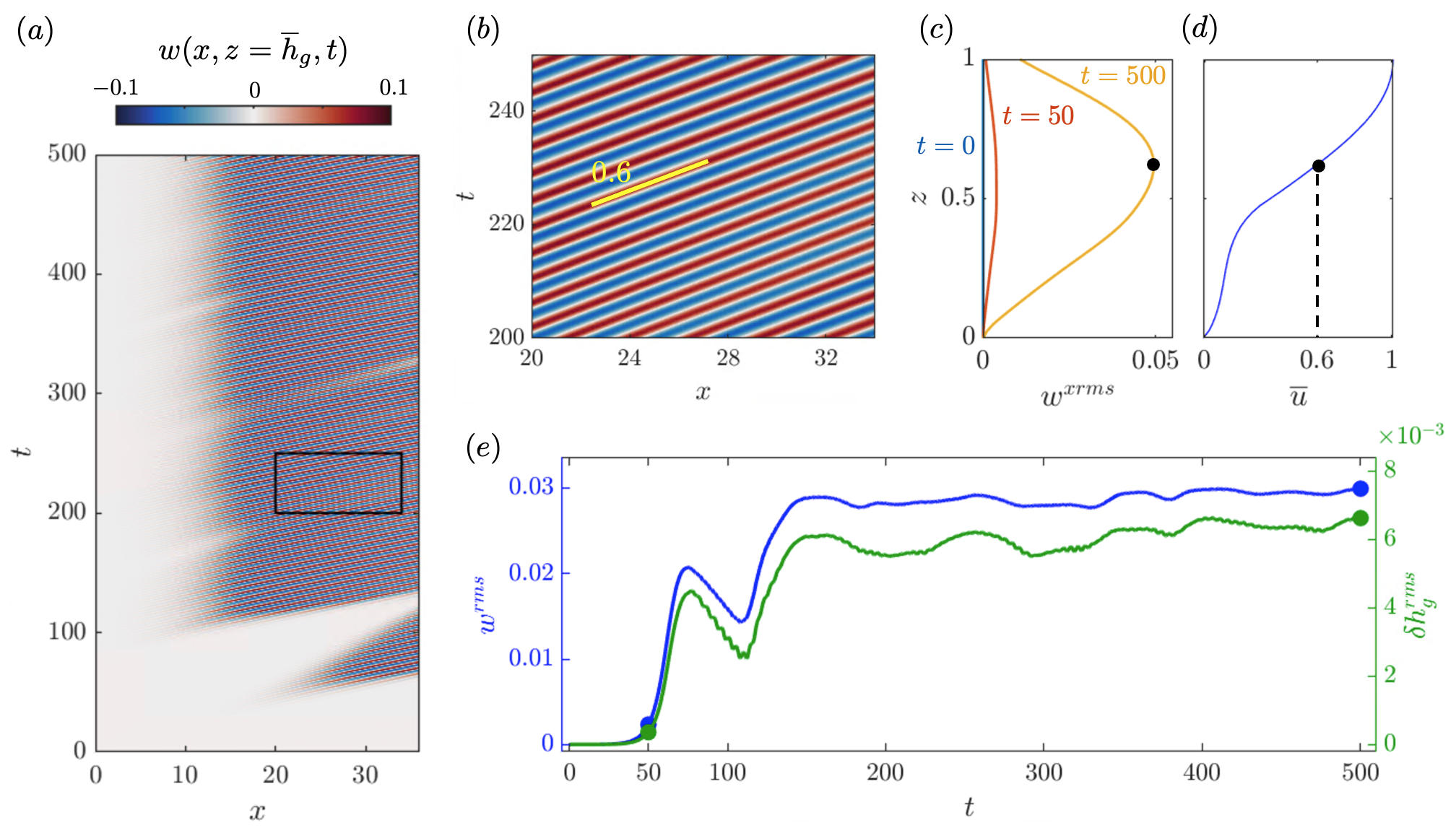}
    \caption{Hovmöler diagrams presenting the space and time evolution of the vertical velocity at the top of the grass bed $w(x,z=\overline{h}_g,t)$ for $(a)$ the full domain and $(b)$ a focused region indicated by the black box in $(a)$.  The yellow line in $(b)$ indicates a constant speed 0.6 of propagation of the perturbations. $(c)$ Vertical velocity rms along $x$ at $t=0$, 50, and 500. $(d)$ Steady-state background horizontal velocity. $(e)$ Time evolution of the spatial rms of $w$ (blue) and $(h_g-\overline{h}_g)$ (green), identifying the instability onset and long-term behavior from both fluid and grass perspectives.}
    \label{fig:result_onsetVisualization}
\end{figure}

The space-time evolution of vertical velocity at the canopy top $z=\overline{h}_g$, where $\overline{h}_g$ is the level corresponding to the steady-state canopy height (Fig.~\ref{fig:result_onsetVisualization}$(a)$), reveals an initial instability onset that originates around $x\approx20$ and $t\approx50$ that is swept out of the domain and replaced by unperturbed flow (video in Supplementary information). Another instability starts at   $t\approx100$, now closer to the inlet at $x\approx 10$, and develops to generate vortices almost periodically. 
Other than small variations that occur over time, notably a weakened vertical velocity field at $x\approx30,\, t\approx300$ and some fluctuation in the onset position, vortices are shed at a regular frequency proportional to the local fluid speed divided by the momentum shear layer thickness~\citep{ho1984perturbed} similar to Kelvin-Helmholtz instability.
The vortices lead to alternating positive and negative $w$ signatures that propagate downstream. 

The generated vortices have constant speed of propagation (0.6), period (4.5), and wavelength (2.7). 
The propagation speed corresponds the mean downstream velocity at the height of the vortex centers. 
The spatial rms of $w$ evaluated along horizontal slices of the domain, $w^{xrms}(z,t)$, at times $t=0$, 50, and 500 (Fig.~\ref{fig:result_onsetVisualization}$(c)$) is used to identify the height of  the maximum $w^{xrms}(z,t=500)$, indicated with a black dot, and corresponds roughly to the height at which the vortex cores propagate (see Fig.~\ref{fig:result_onsetVisualization_field}$(b)$).
In the steady-state horizontal velocity profile (Fig.~\ref{fig:result_onsetVisualization}$(d)$), this height has a horizontal velocity $\overline{u}=0.6$, which matches the speed of propagation of the perturbations  (Fig.~\ref{fig:result_onsetVisualization}$(b)$), that therefore propagate like a convective instability.

The instability onset and strength of vortices can be assessed via the domain-wide rms of the vertical velocity $w^{rms}(t)=   ( \sum_{i,k} w(x_i,z_k,t)^2 / (N_i N_k) )^{1/2}$, where $N_i$ and $N_k$ are the number of grid cells in $x$ and $z$, respectively. 
The vertical velocity rms not only tracks when an instability has developed, but also indicates the strength of the vortices.
The seagrass bed's response to the instability is assessed by the rms value of the grass height perturbation with respect to the steady state height $\overline{h}_g$ over all blade representatives, $\delta h_g^{rms}(t)=(\sum_{i} (h_g(x_i,t) -\overline{h}_g)^2 / N_i )^{1/2}$, and it quantifies the vertical amplitude of grass blade oscillations. 

The two quantities  $w^{rms}$ and $\delta h_g^{rms}$, plotted as a function of $t$ in  Fig.~\ref{fig:result_onsetVisualization}$(e)$, are zero before the onset, when there is no vertical velocity and all blades are at the steady-state shape.  
Their values increase rapidly at the initial onset of instability at $t=50$, they decrease as the initial instability is swept from the domain, then increase again and assume nearly steady values for $t>150$ as the long-term  instability sets in.
Because of the observed plateauing behavior of both curves, long-term values for $w^{rms}$ and $\delta h_g^{rms}$ are defined as the time-average for $t\in[300,500]$ and used to inter-compare different cases in a parametric study.

%%% Subsection %%%

\subsection{Monami kinematics: effect of vortices on seagrass}
\label{subsec:vortex_interaction}

\begin{figure}[t!]
    \centering
    \includegraphics[width=\textwidth]{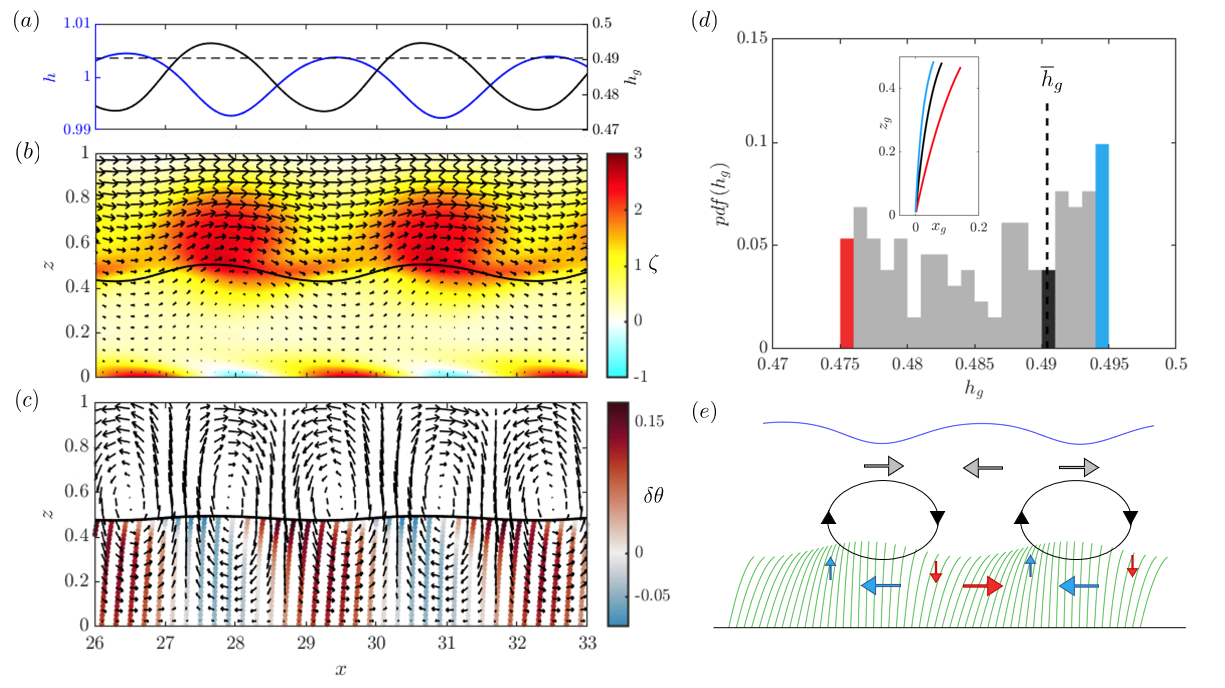}
    \caption{Instantaneous plots at $t=500$, for the $(a)$ surface height $h$ (solid blue line) and grass height $h_g$ (solid black line, with dashed line representing the reference steady-state height $\overline{h}_g$), $(b)$ contours of vorticity $\zeta$ and arrows representing the velocity field $\vect{u}$, with the grass height perturbation (black line) amplified by a factor of 4. $(c)$ Perturbation velocity field $\vect{u}'=\vect{u}-\overline{\vect{u}}$ and grass blade representatives, with colors representing the angle of deflection with respect to the steady-state angle, $\delta \theta = \theta - \overline{\theta}$, in radians.
    $(d)$ Histogram of the grass height distribution within the entire domain, with inset representing the blade shapes corresponding to minimum (red), maximum (blue), and steady-state (black) height (bin width = 0.001). $(e)$ Schematic of how vortices (iso-vorticity contours) induce grass blade deflection, with arrows representing the velocity perturbation induced, and colors indicating how they deflect the grass blade locally. The blue line on top representing the free-surface.}
    \label{fig:result_interaction}
\end{figure}

We analyze the flow field, grass deflections, and free-surface height to evaluate how the instability interacts with the seagrass meadow to produce the oscillatory motion known as monami. 
The shear-driven instability induces a velocity perturbation field $\vect{u}'= \vect{u} - \overline{\vect{u}}$ with respect to the steady-state flow $ \overline{\vect{u}}=(\overline{u}(z),\,0)$ that deflects the grass blades from their steady-state position.

We subsample the domain to visualize two vortices at $t=500$ in Fig.~\ref{fig:result_interaction}$(a-c)$. 
We find that $h(x)$ and $h_g(x)$ are approximately sinusoidal and out of phase, with peaks of $h_g$ slightly lagging troughs of $h$ (Fig.~\ref{fig:result_interaction}$(a)$).
While $h$ has a more symmetric, sine-like profile, $h_g$ is less symmetric, with a steeper increase than decrease.
The vortex cores are centered below the troughs of $h$ and above the peaks of $h_g$ (Fig.~\ref{fig:result_interaction}$(a,b)$).
The grass height $h_g$ is shown in Fig.~\ref{fig:result_interaction}$(b)$ as a thick black line, and we observe that the clockwise vortices induce the grass blades beneath to straighten up. 
Near the seabed, directly below the vortices, negative vorticity values (in blue) indicate that horizontal velocity perturbations are strong enough to reverse the direction of the flow near the no-slip bottom. 
This generates convergence (and divergence) sites along the bed resulting in flow separation points that propagate with the vortices and could export sediment from the seabed. 
The velocity perturbation field calculated with respect to $ \overline{\vect{u}}(z) $  (Fig.~\ref{fig:result_interaction}$(c)$) further helps to visualize the response of grass blades to the flow perturbation.  In Fig.~\ref{fig:result_interaction}$(c)$, blades are colored based on the angle of deviation from the steady-state angle along the blade, $\delta\theta=\theta-\overline{\theta}$.
There are two clockwise eddies that appear in the velocity perturbation field that align with the high vorticity regions in Fig.~\ref{fig:result_interaction}$(b)$.
Additionally, there are counter-clockwise vortices in between the vortex roll up highlighted by the velocity perturbation field, that induce a forward deflection of the blades. 
Blades immediately below the clockwise-vortices straighten up, while blades in between those vortices are subject to the action of counterclockwise-vortices that induce more deformation.

The distribution of $h_g$ (Fig.~\ref{fig:result_interaction}$(d)$) spans $[0.475, 0.495]$  and is asymmetric with respect to the steady-state  $\overline{h}_g$, showing that forward and downward deflection is more common and stronger than upward deflection. This is due to the fact that the drag force that deflects the grass acts normal to the blade. When the grass is downward deflected, the downward vertical velocity helps to enhance the deflection. When the grass is upward deflected with respect to its steady-state shape, the upward velocities are more or less parallel to the grass and do not contribute as much to the grass deflection as the downward velocity.  
The perturbation in the velocity field does not ever reverse the flow in the upper part of the canopy, and as a result the grass blades never move left from the vertical position. 
A schematic of how the vortices induce seagrass motion is presented in Fig.~\ref{fig:result_interaction}$(e)$.
The increased downward deflection of the grass occurs ahead and behind the vortex, where the counter-clockwise perturbation to the mean flow and the downward velocity cause a greater drag on the blades and deflect them forward and downward from their steady-state position (Fig.~\ref{fig:result_interaction}$(c)$).
We identify the sweeps and ejections as corresponding to the perturbed velocity field immediately ahead and behind the vortices, around the seagrass height level (Fig.~\ref{fig:result_interaction}$(b)$). In the sweep region (ahead), the stronger velocity has a downward component and increases downward deflection of the grass. In the ejection region (behind), the weaker velocity has an upward component and induces an upward deflection of the grass.

%%% Subsection %%%
\subsection{Dependence of instability onset and waving amplitude on Reynolds number and grass buoyancy}
\label{subsec:instab_diag}
The shear-driven instability and monami occur only when the drag-induced shear is strong enough for the vortex sheet at the canopy top to become unstable. This occurs when the velocity is large enough, i.e. above some critical value of $Re$, and when the grass-induced drag ($\lambda$) is sufficiently large.
However, we find that $\beta$, the buoyancy parameter, also affects the instability onset and size of vortices.  We use the long-term $w^{rms}$ as an indicator of instability and vortex strength, and $\delta h_g^{rms}$ as an indicator of seagrass waving, to conduct a parametric study in which we vary $Re$ and $\beta$ over a range of values: $Re\in[500,1500]$ and $\beta\in[0.02,0.20]$.   We run a total of $110$ simulations, keeping the other parameters constant: $r=0.5$, $\lambda=1$, and $Fr^2=0.1$. 

\begin{figure}[t!]
    \centering
    \includegraphics[width=.95\textwidth]{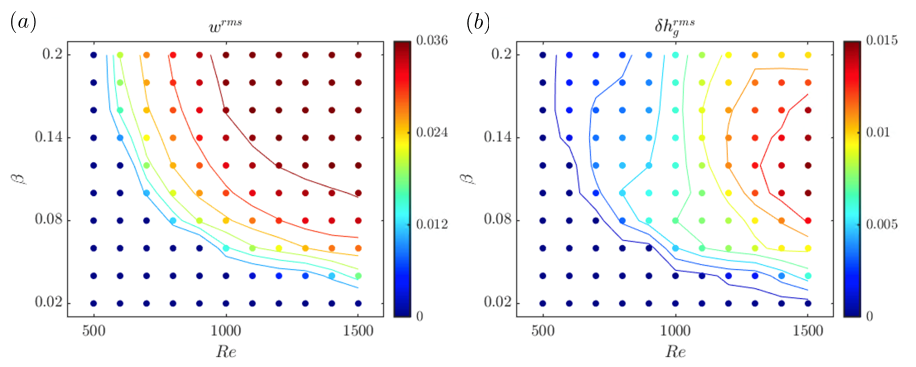}
    \caption{Parametric sweep of $(Re,\beta)$ combinations and the resulting long-term rms of the vertical velocity and grass height perturbation, for $r=0.5$, $\lambda=1$, and $Fr^2=0.1$. Contour plots for $(a)$ $w^{rms}$ tracking instability from the flow perspective, and $(b)$ $\delta h_g^{rms}$ from the seagrass bed perspective. Lines represent contour levels.}
    \label{fig:result_parametric}
\end{figure}

We find no instability ($w^{rms}=0$) for small $Re$ and $\beta$ and that $w^{rms}$ increases for increasing values of  $Re$ and $\beta$, saturating for high values of both parameters (Fig.~\ref{fig:result_parametric}$(a)$). 
Larger $w^{rms}$ corresponds to stronger vortices inducing vertical velocities of greater magnitude, and it is reasonable to expect that as $Re$ increases for fixed $\beta$ the shear layer becomes stronger as do the resulting vortices.
As $\beta$ is increased (for fixed $Re$), the more buoyant blades are less deflected from the vertical.  This sharpens the mean velocity gradient ($\dv*{\overline{u}}{z}$) and results in stronger instabilities. 
Less buoyant blades are more deflected for the same $Re$, impose less drag (which is largely due to the velocity component normal to the grass) and inhibit the development of instabilities for fixed $Re$. Though our model uses buoyancy, the result is that the greater the deflection of stems, the less strong the instability, regardless of whether the deflection results from weak bending stiffness or buoyancy. 

The critical combinations $(Re,\beta)$ in the instability diagram above which $w^{rms}>0$ define an instability curve in Fig.~\ref{fig:result_parametric}$(a)$. This curve is in agreement with the result that shear at the top of the canopy is the relevant criterion in determining the stability of steady unidirectional flows~\citep{wong2020shear}, as the velocity shear magnitude grows with both $Re$ and $\beta$.

While $w^{rms}$ grows monotonically with $Re$ and $\beta$ (Fig.~\ref{fig:result_parametric}$(a)$), a non-monotonic behavior of the amplitude of waving, assessed by $\delta h_g^{rms}$, is observed in Fig.~\ref{fig:result_parametric}$(b)$. 
In general, the amplitude of grass motion or $\delta h_g^{rms}$ increases with $Re$. However, the maximum $\delta h_g^{rms}$ occurs for intermediate values of $\beta$. Small $\beta$ suppresses the shear instability and creates small $w^{rms}$, whereas for larger values of  $\beta$ (e.g. $0.20$), the buoyancy of the grass resists its deformation, even though the fluid instability and $w^{rms}$ are stronger. 
For large $\beta$, the grass is almost vertical and the vortices do not induce an observable oscillatory motion. 
Grass oscillations are therefore maximized for specific combinations of $(Re,\beta)$, with maxima $\delta h_g^{rms}$ observed for high $Re$ with intermediate $\beta$ values.
The observed trends point to the fact that experiments using rigid dowels~\citep{ghisalberti2002mixing,ghisalberti2006structure} may overestimate the strength of the induced vortices, or not accurately predict their development, compared to what would be observed for more realistic, deformable seagrass beds.

\begin{figure}[t!]
    \centering
    \includegraphics[width=\textwidth]{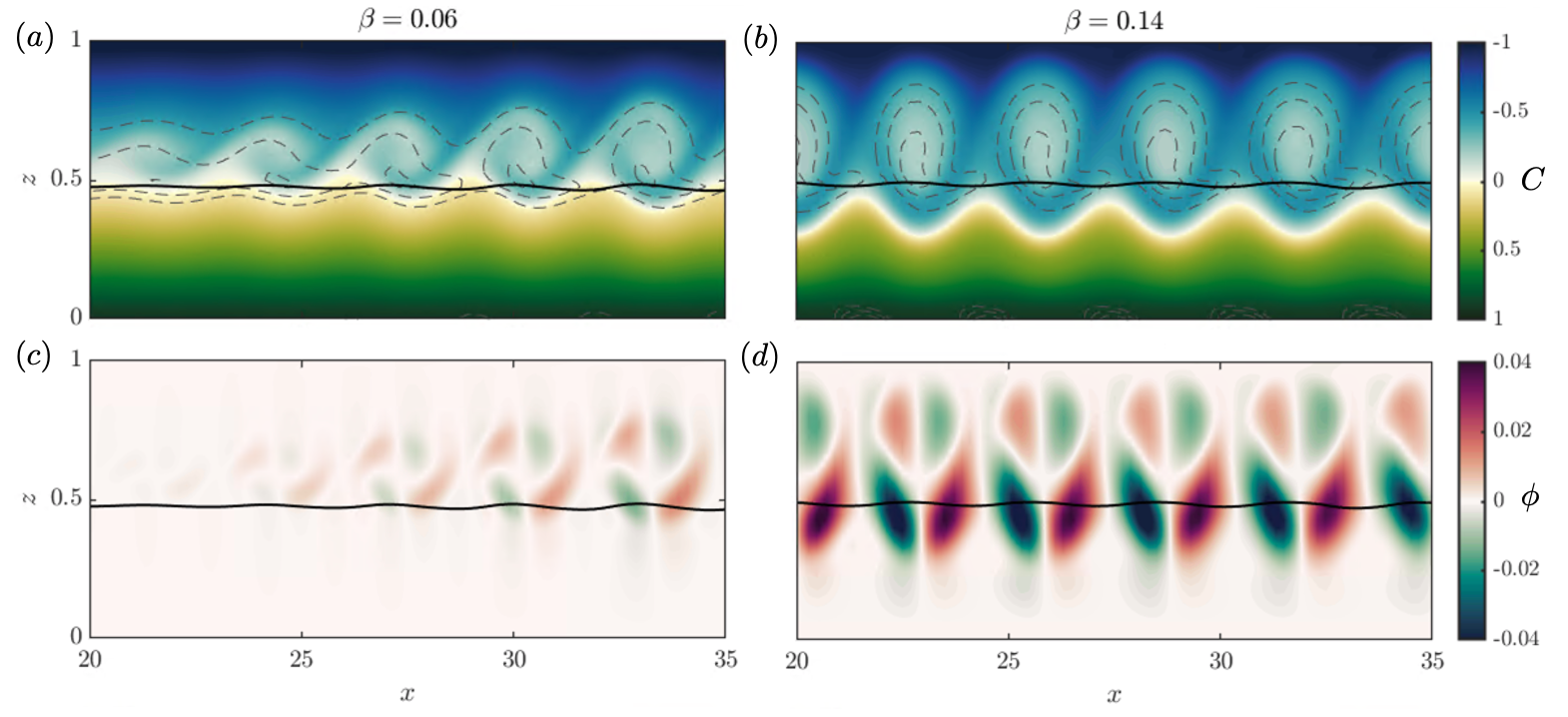}
    \caption{Vortex size and tracer transport variability for $\beta=0.06$ and $0.14$. Instantaneous plots of $(a,b)$ the tracer concentration $C$, with gray dashed lines representing iso-vorticity contours ($\zeta=1.5,\,2,\,2.5$), and $(c,d)$ vertical tracer flux $\phi$, at $t=500$, with $Re=1000$ and $\beta=0.06$ and $\beta=0.14$, respectively. The solid black line is the seagrass height $h_g$.
    }
    \label{fig:result_betavariation}
\end{figure}

\begin{figure}[t!]
    \centering
    \includegraphics[width=\textwidth]{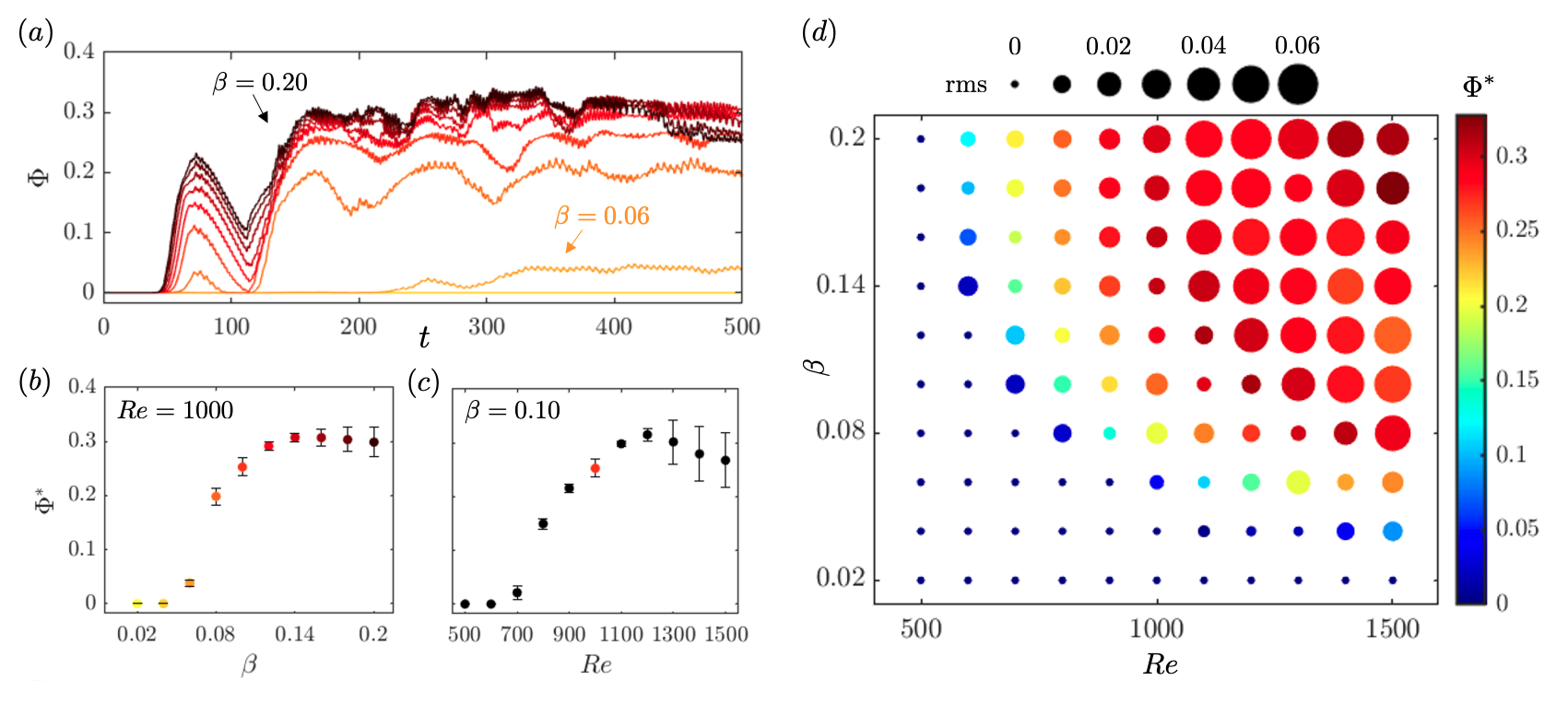}
    \caption{$(a)$ Tracer exchange $\Phi$ as a function of time for $\beta\in\{0.02, 0.04,\ldots, 0.20\}$. Long-term exchange $\Phi^*$ $(b)$ as a function of $\beta$, for $Re=1000$, $(c)$ as a function of $Re$, for $\beta=0.10$, and $(d)$ as a function of $(Re,\beta)$. Error bars in $(b,c)$ correspond to the rms deviation from the mean, and colors match curves in $(a)$. In $(d)$, the colors represent the mean value $\Phi^*$ for the given parameter combination, and the marker size represents the rms deviation from the mean.
    }
    \label{fig:result_exchange_grid}
\end{figure}

%%% Subsection %%%
\subsection{Material exchange across the grass bed}
\label{subsec:tracer_transport}

To evaluate the impact of the vortices and the grass deflection on material exchange between the seagrass bed and the overflow, we model a tracer field and evaluate its transport.
The initial tracer distribution $C_0(z)= 2\left(\overline{h}_g - z\right)$ is linear in $z$ with $C_0=0$  at the canopy top $z=\overline{h}_g$ , $C_0>0$ within the grass bed and $C_0<0$ in the overflow. 

The tracer transport and exchange is assessed for the same range of $Re$ and $\beta$ as above. 
A snapshot of the tracer distribution at $t=500$ highlights how the material transport resulting from the vortices changes with grass buoyancy for $\beta=0.06$ and $\beta=0.14$ in Fig.~\ref{fig:result_betavariation}$(a,b)$.
The cores of the vortices lie predominantly above the canopy and the majority of material entrained from the seagrass bed by the vortices appears to come from the upper region of the grass bed (Fig.~\ref{fig:result_betavariation}$(b)$)
despite the vortex velocity signature extending well into the grass. Iso-vorticity contours highlight the alignment of the material vortices (Figs~\ref{fig:result_betavariation}$(a,b)$) and the cores of highest vorticity.

Less buoyant blades, $\beta=0.06$, in Fig.~\ref{fig:result_betavariation}$(a)$  allow for a larger mean grass deformation and smaller vortices.
As the vortices propagate down the channel in Fig.~\ref{fig:result_betavariation}$(a)$, they grow in size and in the amount of material entrained. At the same time, material from the overflow is entrained into the grass at a greater rate as well.
Vortices are shed from a more or less fixed location $x\approx15$ and grow primarily in vertical extent, as the wavelengths appear to be approximately constant.
For $\beta=0.14$ in Fig.~\ref{fig:result_betavariation}$(b)$, with more buoyant blades, the vortex size is effectively constant throughout the same domain, meaning that the development region from no vortex to fully developed vortex is shorter compared to the previous case.

To quantify the vertical flux of tracer $ \phi(x,z,t) $ and the tracer exchange induced by these instabilities, we define the vertical tracer flux  as the product of the local vertical velocity and the tracer perturbation with respect to $C_0$, expressed as
$\phi = w\,C'$, where $C' = C(x,z,t)-C_0(z)$.
The instantaneous vertical flux at $t=500$ in Figs~\ref{fig:result_betavariation}$(c,d)$ for $\beta=0.06$ and $\beta=0.14$, corresponding to the tracer fields in Figs~\ref{fig:result_betavariation}$(a,b)$, is positive or negative depending on the co-variance between $w$ and $C'$; positive flux results from the upward (and downward) movement of anomalously high (and low) tracer anomaly.  The flux both out of, and into, the grass grows as the vortex propagates down the channel for smaller $\beta$ (Fig.~\ref{fig:result_betavariation}$(c)$) .  
While an individual vortex is experiencing progressively more exchange as it propagates down the channel, the domain as a whole has reached steady-state with regards to the amount of exchange taking place.
For $\beta=0.14$ (Fig.~\ref{fig:result_betavariation}$(d)$) the vortices are fully developed and have a constant size and the exchange into and out of the grass bed is balanced. The overturning that occurs inside the vortex cores is reflected by the red-green lobe patterns where the values of $\phi$ are dominated by the vertical velocities. 

To quantify the relative amount of tracer exchange occurring between the seagrass domain and the overflow, we define the tracer exchange $\Phi$ at $z=\overline{h}_g$, for $x\in[x_a, x_b]$, as
\begin{linenomath*}
\begin{align}
    \Phi(t) = \int_{x_a}^{x_b}|\phi(x,z=\overline{h}_g,t)|\,{\rm d}x.
\end{align}
\end{linenomath*}
Both $\phi$ and $\Phi$ should be viewed as relative, as their value depends on the initial tracer distribution. The time evolution of $\Phi(t)$ is plotted in Fig.~\ref{fig:result_exchange_grid}$(a)$ for $\beta\in[0.02,0.20]$, with $x_a=20$ and $x_b=35$.
The small oscillations observed in each of these curves relates to the vortex turnover time ($\approx 10$) and to new vortices entering and leaving the domain of integration (see Supplementary information for a video showing how the flux field $\phi$ synchronizes with the time-evolution of $\Phi$ for $\beta=0.06$ and $\beta=0.14$).
Focusing of the long term variations, we observe that the exchange $\Phi$ grows once the instability starts, and plateaus in all cases for $t>300$. 
We therefore define the long-term average exchange $\Phi^*$ as the time-average of $\Phi(t)$ for $t\in[300, 500]$.
Tracer exchange, measured by this metric, is higher for less deformable blades (larger $\beta$), when all other parameters kept constant.
$\Phi^*$ varies with the buoyancy parameter $\beta$ and with $Re$ (Figs~\ref{fig:result_exchange_grid}$(b,c)$). 
The increase in exchange with $\beta$ and $Re$ eventually saturates for  $\beta=0.10$ and $Re\geq1300$. For larger $Re$, we observed a slight decrease in $\Phi^*$ alongside an increase in the uncertainty associated with the long-term rms value.

The parametric study previously presented is used to investigate the dependence of the long-term average exchange $\Phi^*$ on ($Re$, $\beta$). The exchange in Fig.~\ref{fig:result_exchange_grid}$(d)$ shows strong correlation with the $w^{rms}$ trends previously observed in Fig.~\ref{fig:result_parametric}$(a)$. This result is in agreement with the comment in~\citet{nepf2008flow} that the exchange of a scalar would follow the same trend of the exchange of momentum, and decrease as canopy deformability and motion increases. For growing $Re$ or $\beta$, however, larger uncertainty is observed. This is also seen in Figs.~\ref{fig:result_exchange_grid}$(b,c)$ for large $\beta$ and $Re$, respectively, and is potentially related to vortex merger events that become more common with increasing $Re$ and induce temporal fluctuations on the exchange (Supplementary information).

%%%%%%%%%%%%%%%%%%%%%%%%%%%%%%%%%%%%%%%%%%%%%%%%%%%%%%%%%%
% Discussion

\section{Discussion}
\label{sec:Dircussion}

% How our results link to previous studies
In our model, where buoyancy and fluid drag determine the grass blade shape, we find that the  Kelvin-Helmholtz-like flow instability weakens with more deflection of the blades. We hypothesize that this result will hold regardless of whether the grass deflection is restored by bending stiffness or buoyancy. As our blades stand up by buoyancy and have no flexural rigidity, the buoyancy parameter $\beta$ controls the degree of deformation of the grass blades and yields analogous results to the Cauchy number~\citep{luhar2016wave,wong2020shear} for flexible beam models (Supplemental information).  This hypothesis is further supported by similarities between experiments featuring flexible blades with known elasticity~\citep{ghisalberti2006structure} and our model. Changing the deflection mechanism, therefore, should not impact the qualitative behavior.

Schematics from previous studies indicated that clockwise-vortices forward-deflect the grass blades immediately below~\citep{ghisalberti2002mixing,nepf2008flow,nepf2012flow,okamoto2016flow,wong2020shear}. We find to the contrary that the vortices straighten the grass directly below their core because the horizontal velocity perturbation induced by the vortices acts in the opposite direction to the mean flow. This results in a lower drag force on the grass below the vortex core relative to steady-state, which makes the grass blades more erect. 
Our schematic of how the vortices induce seagrass motion (Fig.~\ref{fig:result_interaction}$(e)$) provides a correction to previous schematics in the literature and is consistent with the experimental measurements in~\citet{ghisalberti2006structure} (figures 7 and 8), which report that the smallest horizontal velocity aligns with where the grass blades are most erect and the highest velocities where the grass is most deflected.
% Our velocity perturbations have similar magnitudes in the $x$ and $z$ directions, while in~\citet{ghisalberti2006structure} the vertical velocity perturbations are an order of magnitude smaller than the horizontal perturbations.
% This is likely due to the larger drag impedance $\lambda$ in our simulations compared to the laboratory experiment, so that our seagrass bed has a stronger effect in the flow and induces stronger velocity perturbations relative to the mean. 
Our result that more deformable seagrass blades inhibit tracer exchange is consistent with the observations. \citet{nepf2008flow} find that exchange of momentum is most efficient for rigid canopies and exchange efficiency decreases as the deflection of the canopy increases. They conjecture that the exchange of a scalar should follow a similar trend and decrease as canopy deformability and deflection increase.

% New studies that would be of interest based on the results of our model
Our model of the instability and seagrass also highlights some phenomena within the canopy  that warrant further study.
For cases of large $\lambda$, where the flow speed within the canopy is much smaller than the overflow, the velocity perturbations induced by the vortices cause flow reversal near the seabed, which results in flow separation points that propagate with the vortices and could export sediment from the seabed.
Sediment resuspension related to the presence of seagrass canopies has been observed in the field~\citep{adams2016feedback} and quantified in laboratory experiments, where canopies were found to increase seabed sedimentation compared to bare substrates, and the more blades per unit area, the greater the amount of sediment deposited on the seabed~\citep{barcelona2021particle}.
Another feature that is considered in our model is the free surface and its variation relative to the position to the vortex.
While~\citet{mandel2019surface} have measured experimentally the surface signature of the shear-instability that develops as flow moves through a canopy of rigid rods, their study does not consider the relative phase of the surface signature and the induced vortices.
However, they provide a schematic indicating the wave crests immediately above the vortices, which is $180\degree$ out of phase with our model. 
Further experimental investigation of the relative phase as well as an analysis of the free-surface signature of monami with a moving grass bed could provide insight into the potential of remotely observing monami. 

% Future improvements of the model
Despite qualitative agreement with experiments that feature larger scale oscillations of the grass, our model produces small amplitude blade oscillations. 
Higher order effects that have been neglected in our model, such as grass inertia, added mass, and virtual buoyancy~\citep{luhar2016wave, wong2020shear}, can be incorporated to more accurately model the grass meadow for oscillations of higher amplitude. 
Inertia may introduce another characteristic frequency to the oscillatory motion, and tracking how the blade moves in time and using the instant relative velocity between fluid and blade would more accurately represent the drag for faster, higher amplitude grass motion.
Further refinement of the method used to distribute the blade forces onto the computational grid to account for the position of the blade and the center of neighboring cells in both $x$ and $z$ direction is also desirable, as it allows for a more accurate description for higher amplitude oscillations. 

% Future studies with the model
While we focused exclusively on the effects of $Re$ and $\beta$ in this study, variations of the dimensionless parameters $r$, $\lambda$, and $Fr$ also impact the results and can be addressed using the current version of the model. 
Additionally, the model can handle variations in the canopy to free-surface height ratio $r$ and spatially uneven $\lambda(x,\,z)$. It can be used to study variable blade number per unit area $\overline{N}$, spatially variable grass parameters $b, d$, and drag coefficient $c_D$.
In the current study we explored the two-dimensional vortex regime, but performing three-dimensional simulations with the model would be especially beneficial to study instabilities along the $y$-direction and vortex interactions. 
Our model could also be used to compare with field studies of seagrass meadows that quantify fluid exchange above and within the canopy~\citep{hansen2017turbulent}.
Additionally, a study of tracer and sediment transport could be undertaken from a Lagrangian perspective, by applying coherent structure detection methods. This could be used to more conclusively explain how the grass motion impacts material exchange.
Finally, applications are not constrained to aquatic vegetation, and our model can simulate atmospheric flows through forests by adjusting the dimensionless parameters to produce the typical canopy deformations and velocity magnitudes.

%%%%%%%%%%%%%%%%%%%%%%%%%%%%%%%%%%%%%%%%%%%%%%%%%%%%%%%%%%
% Conclusion
\section{Conclusions}
\label{sec:Conclusion}

Our two-phase model of buoyant, deformable, non-shear resistant seagrass blades
captures the interaction of flow and submerged canopies, yields shear-instabilities that evolve into vortices and induce an oscillatory motion of the grass blades.  
While previous schematics of the vortex-grass interaction feature the greatest deflection of the grass immediately below the vortices, our model demonstrates that the velocity perturbation induced by these clockwise vortices acts to make the grass immediately below the vortex more erect than the surrounding canopy, forming the maxima in canopy height. Perturbations induced by the vortices to the background flow increase deflection ahead and behind the vortex.

A stability study of the system as a function of $(Re,\beta)$ demonstrates the onset of instability.
The vertical velocity induced by the instability increases with both $Re$ and $\beta$.
As $Re$ increases, the shear layer strength increases resulting in stronger vortices.
Increased $\beta$, corresponding to more buoyant grass that is less deformable, also produces stronger vortices, indicating that the deformability of the grass reduces the vortex strength and delays the instability onset.
A scalar field advected with the flow is used to quantify material exchange between seagrass and overflow.
Tracer exchange is a function of $Re$ and $\beta$, with less deformable blades or larger $Re$ leading to an increased shear above the canopy that results in stronger vortices inducing more exchange.
Grass deformation, therefore, inhibits fluid exchange by decreasing the shear magnitude at the canopy top, and therefore the resulting vortex sizes and induced vertical velocities.

%%%%%%%%%%%%%%%%%%%%%%%%%%%%%%%%%%%%%%%%%%
% Methods section (in manuscript):
\section{Methods}
\label{sec:Methods}

\subsection{Non-hydrostatic model formulation}
\label{subsec:model_formulation}

Numerical simulations are run using a version of the Process Study Ocean Model (PSOM)~\citep{mahadevan1996a_nonhydrostatic,mahadevan1996b_nonhydrostatic}, a finite-volume, non-hydrostatic model that we modify to account for the seagrass drag, recompute the blade shapes at each time step as a function of the velocity field, and solve the two-way coupled system of equations in dimensionless form.
The governing equations for the fluid are the incompressible Navier-Stokes equations with an added body force term that models the seagrass drag on the fluid and is exclusively applied within the seagrass phase.  
Following the process of homogenization in~\citet{wong2020shear}, this term accounts for the effects on the flow from all resulting forces applied by multiple seagrass blades on the fluid.  
The free-surface $h$ satisfies the integral form of the kinematic condition, and a scalar tracer field of concentration $C$ is advected with the flow.  

We consider a dimensionless system of equations similar to the one used by~\citet{wong2020shear}, with a critical difference being the inclusion of the buoyancy parameter $\beta$ to restore the grass deflection instead of the Cauchy number for bending stiffness.  
The five dimensionless parameters $Re$, $\beta$, $r$, $\lambda$, and $Fr$ defined in (\ref{eq:nd-parameters}) uniquely determine the flow characteristics and can be varied independently by tuning the dimensional parameters $\nu^*$, $d$, $\ell$, and $\overline{N}$, respectively.
The resulting dimensionless governing equations, obtained using the characteristic scales $[x,\,z,\,h,\,h_g]=H$, $[\mathbf{u}]=U$, $[t]=H/U$, $[p]=\rho U^2$,  and $[C,\,\theta]=1$, are
\begin{linenomath*}
\begin{align}
    \boldnabla\cdot\vect{u}=0, \quad
    \md{\vect{u}}{t} + \frac{1}{Fr^2} \boldnabla_H h + \boldnabla{p} - \frac{1}{Re}\nabla^2\vect{u} + \lambda \sec\theta\,\vect{f} = 0, \label{eq:ns-momentum-vec-dimensionless} \quad
    \pd{h}t + \boldnabla_H\cdot\int_{0}^{h}\vect{u}\, \dd z = 0, \quad
    \md{C}{t} = 0. 
\end{align}
\end{linenomath*}

For all the simulations presented, the steady-state velocity profile $\overline{u}(z)$ is imposed at both the inlet and outlet, and a restoring term is applied to a buffer region at the outflow boundary (omitted in Fig.~\ref{fig:full_domain_intro}) for $x\in[L,\,L_r]$.  The buffer is used to suppress the vortices so the flow matches the outflow boundary condition, thereby minimizing reflections from the boundary. The bottom boundary has a no-slip condition for the velocity ($u=w=0$). Pressure is constant at the free-surface.
In this study, we explore two-dimensional solutions by allowing no variations in the $y$-direction within the model.

\subsection{Buoyant blade equations} \label{subsec:grass_equations}

\begin{figure}[t]
    \centering
    \includegraphics[height=15em]{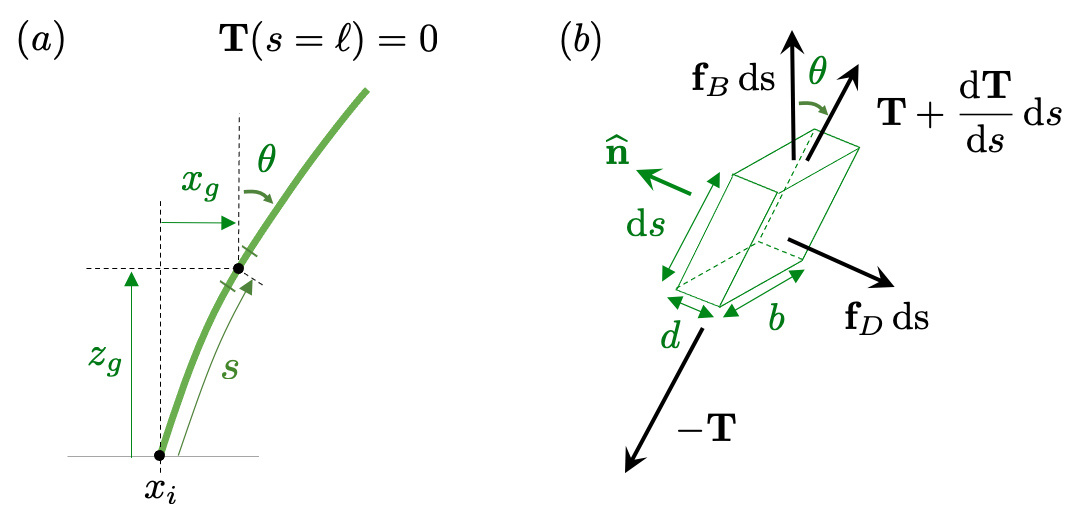}
    \caption{$(a)$ Schematic of a grass blade representative, rooted at $x_i$, with grass coordinates in green. $(b)$ Forces acting on an infinitesimal, inextensible blade element: tension $\mathbf{T}$, drag per unit length $\mathbf{f}_D$, and buoyancy per unit length $\mathbf{f}_B$. }
    \label{fig:blade_schematic}
\end{figure}

The seagrass bed is modeled using a single seagrass blade representative per cell center used in the numerical simulations (with the $i$-th representative rooted at $(x=x_i,\,z=0)$, Fig.~\ref{fig:blade_schematic}$(a)$).
There are $\overline{N}$ grass blades per unit area, and each representative models the local averaged blade shape and contribution to the flow. 
Neglecting flow in the $y$ direction, the motion of each blade representative is confined to the the $xz$ plane, and blade representatives are uniformly distributed in the $x$-direction.

All blade representatives are inextensible and have constant length $\ell$. We solve for their shape as a function of the grass buoyancy and fluid load due to drag.
The shape of a blade is described by the coordinates $\vect{x}_g(s)=(x_g(s),\,z_g(s))$, which are measured with respect to its base at $x=x_i$ and parameterized by the distance $s$ along the blade ($s=0$ at the base and $s=\ell$ at the tip, Fig.~\ref{fig:blade_schematic}$(a)$).
The blade coordinates in the $xz$ plane are uniquely determined by $x_i$ and the clockwise blade angle $\theta(s)$ with the vertical.
The blade has width $b$ (along $y$) and thickness $d$ (along $x$).

At every instant $t$, we assume
%operate under the assumptions that $(i)$ the grass motion is slow enough compared to the flow speeds that the relative velocity between blade and flow is approximately the flow velocity; $(ii)$
the grass blades are in equilibrium with the flow, which corresponds to neglecting the blade inertia (note that typically $d\ll b$ and the blade acceleration is small, which makes the inertial term negligible compared to the drag and tension contributions). The blades are buoyant ($\rho_g<\rho$) and do not resist shear (their flexural rigidity $EI$ is negligible) %, as $I\propto bd^3$). 
Under these assumptions, the only three forces acting on the blade are: tension, drag, and buoyancy (Fig.~\ref{fig:blade_schematic}$(b)$).
Other force terms affecting the blade motion~\citep{luhar2016wave}, such as virtual buoyancy and added mass, are neglected under our assumptions.

The tension $\mathbf{T}=(T^x, T^z)$ is oriented along the blade. We assume that drag acting tangential to the blade is negligible. 
The drag per unit length $\mathbf{f}_D$ is normal to the blade and obeys a quadratic drag law, and the buoyancy per unit length $\mathbf{f}_B$ points upward:
\begin{linenomath*}
\begin{align}
    \label{eq:f_grass}
    \mathbf{f}_D = -\frac12\rho b c_D (\mathbf{u}\cdot \unitvect{n})|\mathbf{u}\cdot \unitvect{n}|\,\unitvect{n}, \quad \mathbf{f}_B = (\rho-\rho_g)gbd\, \unitvect{z},
\end{align}
\end{linenomath*}
where $\unitvect{n}=(-\cos\theta,\,\sin\theta)$ is the upstream normal vector to the blade and $\unitvect{z}=(0,\,1)$ is the unit vector pointing upward.  

After non-dimensionalizing the tension, drag, and buoyancy, the force balance for the blade element becomes
\begin{linenomath*}
\begin{align}
    \td{\vect{T}}{s} + \vect{f} + \beta \unitvect{z} = 0, \label{eq:T_vec} \quad \textrm{where} \quad \mathbf{f} = -\frac12 (\mathbf{u}\cdot \unitvect{n})|\mathbf{u}\cdot \unitvect{n}|\,\unitvect{n}.
\end{align}
\end{linenomath*}
The boundary condition $\vect{T}(s=\ell)=\mathbf{0}$ at the grass tip allows us to solve for $\vect{T}(s)$ by integrating \eqref{eq:T_vec} from tip to base.
Because $\mathbf{T}\cdot\unitvect{n}=0$, the local blade angle is
\begin{linenomath*}
\begin{align}
    \theta = \tan^{-1}\left(\frac{T^x}{T^z}\right),
\end{align}
\end{linenomath*}
under the assumption that the blade does not overturn ($T^z>0$ and $|\theta|<\pi/2$ along the blade).
Finally, integrating
\begin{linenomath*}
\begin{align}
    \td{x_g}{s}=\sin\theta, \quad \td{z_g}{s}=\cos\theta
\end{align}
\end{linenomath*}
from the root up (for $s=0$ to $\ell$) uniquely determines the instantaneous blade shape if the velocity field is known.

\subsection{Fluid-blade coupling} \label{subsec:coupling_equations}

The two-way coupling in our model accounts for the impact of the fluid velocity on the shape of the grass, as well as that of the grass shape on the fluid, via the drag force.  
The canopy height $h_g(x,t)$ separating the two phases -- seagrass and overflow -- depends on the instantaneous positions of all blades. 
Once the coordinates for all blade tips at the instant $t$ have been determined, $h_g(x,t)$ is obtained by fitting a spline through the blade tips.

The relationship between the drag force per unit length $\mathbf{f}$, that acts on a blade representative and is used to solve for the blade shape, and the drag force per unit volume $\mathbf{F}$, that acts on any point in the fluid and appears in the momentum equations, is obtained through the process of homogenization presented in~\citet{wong2020shear}. The force $\mathbf{F}$ on the fluid is proportional to $\lambda$ and to the secant of the local blade representative angle $\theta$, which physically accounts for an increase in the effective number of blades per unit area when neighboring plants tilt. Everywhere within the grass bed, where $z \leq h_g(x,t)$, $\mathbf{F} = - \lambda\sec\theta \,\,\mathbf{f}$, and at the overflow, where $z > h_g(x,t)$, we set $\mathbf{F}=\mathbf{0}$ .

Note that while $\mathbf{F}$ is defined at any location $(x,z)$, $\mathbf{f}$ and $\theta$ are evaluated along the blade representatives only and are a function of $(i,s)$.
In order to distribute the drag from each blade element to its neighboring cells, we use a Gaussian kernel in the $x$-direction, with standard deviation $0.3\Delta x$.
The relationship between $\mathbf{f}$ and $\mathbf{F}$ couples the equations for fluid and grass blades, and an iterative under-relaxation method is used at each time step for each grass representative to attain convergence to the equilibrated grass shape.

\subsection{Numerical grid and conformal map}
\label{subsec:numerics_fluid}

A conformal map for the vertical grid coordinate accounts for variations in time of the seagrass height $h_g(x,t)$ and free-surface height $h(x,t)$, allowing us to reproduce the monami dynamics in an open channel while maintaining a uniform grid in the transformed space.  This requires transforming the equations and boundary conditions to solve them in the computational domain, as time and the physical domain evolve~\citep{mahadevan1996a_nonhydrostatic,mahadevan1996b_nonhydrostatic}.

We discretize the physical domain with a smooth boundary fitted curvilinear grid, and map this domain onto a computational grid that is rectangular, uniform, and has $N_i$ by $N_k$ grid intervals in the $x$ and $z$ directions, respectively.
The free-surface ($z=h(x,t)$) is mapped onto the top boundary of the rectangle in the computational domain. The seabed ($z=0$) is mapped onto the bottom boundary of the rectangle. Finally, the top of the seagrass bed ($z=h_g(x,t)$) is mapped to the top edge of the $N_{kg}$-th cell row in the computational domain, so that the bottom $N_{kg}$ cell layers correspond to the seagrass phase, where drag is applied, and the top $(N_{k} - N_{kg})$ cells correspond to the overflow phase, where there is no drag. Cells corresponding to each of the two phases are illustrated in green and blue in Fig.~\ref{fig:full_domain_intro}.

The simulations presented here use $N_i=216$, $N_k=48$, and $N_{kg}=24$. The grid spacing in the physical domain is $\Delta x=0.2$ horizontally, so that $L_r=43.2$ (the physical channel is 43.2 by 1), and $\Delta z\approx0.02$ vertically (note that $\Delta z$ is non-uniform and varies depending on $h_g$ and $h$), and a uniform time step $\Delta t=0.1$ is used. The horizontal length of the domain before the buffer region where the velocity profile is restored to $\overline{u}(z)$ is $L=36$, and the dimensionless domain considered for the analysis is $x\in[0,L]$, $z\in[0,1]$.

%%%%%%%%%%%%%%%%%%%%%%%%%%%%%%%%%%%%%%%%%%%%%%%%%%%%%%%%%%%%%%%%%%%%%%%%%%%%%%%%
%%% REFERENCES
%%%%%%%%%%%%%%%%%%%%%%%%%%%%%%%%%%%%%%%%%%%%%%%%%%%%%%%%%%%%%%%%%%%%%%%%%%%%%%%%
% \clearpage
% \nocite{*} % Comment this line to show only the ones cited in the text
\bibliography{references.bib}

\appendix
\renewcommand{\thefigure}{S\arabic{figure}}
\setcounter{figure}{0}
\nolinenumbers

\newpage
% \section{Supplementary Information A}
% \label{sec:Supplementary_Material}

% \small{
% \noindent \emph{Seagrass deformation affects fluid instability and tracer exchange in canopy flow (Vieira, Allshouse \& Mahadevan, 2022)}
% }

\clearpage
\section{Comparison between buoyancy and rigidity models}
\label{subsec:buoyancy_vs_rigidity}
\thispagestyle{empty}

Prior to the onset of instability, or closer to the inflow section of the channel, the flow and grass are both steady.  The steady state solution is a function of $z$ alone, and can be calculated with a simplified one-dimensional coupled model that eliminates dependence in $x$ and $t$ (Methods Section). 

\begin{figure}[h!]
    \centering
    \includegraphics[width=0.65\textwidth]{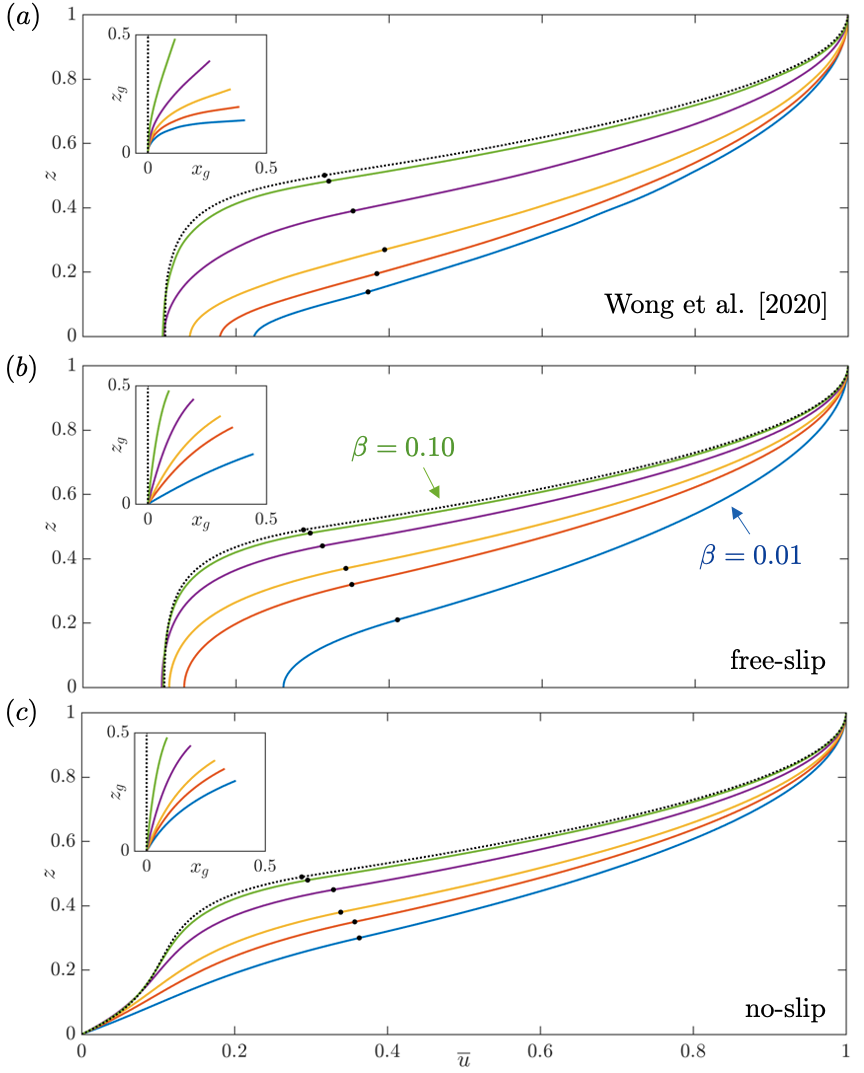}
    \caption{Steady-state horizontal velocity profile $\overline{u}(z)$ and corresponding blade configurations (inset), for $Re=10^3$, $r=0.5$, and $\lambda=1$. $(a)$ Results from \citet{wong2020shear}, flexible grass blade, with $C_Y/Re^2=10^{-1}$, $10^{-1.5}$, $10^{-2}$, $10^{-2.5}$, $10^{-3}$, and 0 (dotted). Buoyant grass model, using a $(b)$ free-slip bottom and $(c)$ no-slip bottom boundary condition, and $\beta=0.01$, 0.015, 0.02, 0.04, 0.10, and $\infty$ (dotted lines). Same $\beta$ values are plotted using the same colors in $(b,c)$. The black dots on the velocity profiles mark the position of the corresponding canopy height $\overline{h}_g$ in each case.}
    \label{fig:single_stem}
\end{figure}

The steady state velocity profiles $\overline{u}(z)$ and corresponding grass positions $(\overline{x}_g, \overline{z}_g)$ calculated for $Re=10^3, r=0.5, \lambda= 1$, $Fr^2=0.1$, are presented for a range of values of the buoyancy parameter $\beta$ (Fig.~\ref{fig:single_stem}). 
Solutions are computed for two different fluid boundary conditions:   $\dv*{\overline{u}}{z}=0$ (free-slip) at the bottom ($z=0$) (Fig.~\ref{fig:single_stem}$(b)$) and $\overline{u}=0$ (no-slip) (Fig.~\ref{fig:single_stem}$(c)$). At the surface  ($z=1$),  $\dv*{\overline{u}}{z}=0$. 

%To study the strength of the resulting vortices, it is valuable to model how the instabilities arise from the steady state velocity and grass shape.
%Our first step is to calculate the steady state velocity profile $\overline{u}(z)$ and the resulting grass deflection for a given set of dimensionless parameters, while varying the buoyancy parameter $\beta$.
%The calculation of the steady state velocity profile and grass shape also provides a means for comparison of our buoyancy dominant model with the flexural rigidity dominant model developed of \citet{wong2020shear}.

%To obtain the steady state profiles, we set variability with time and the along flow coordinate $x$ to zero in the model.  
%The steady-state shape for a single blade (or many blade representatives in a seagrass bed subject to a velocity profile that depends only on $z$) sets $\overline{\theta}(z)$ and $\overline{h}_g$.
%The overbar is used hereafter to represent the steady-state solutions, which are independent of $x$ and $t$.
%The fluid boundary conditions are either $\overline{u}=0$ (no-slip) or $\dv*{\overline{u}}{z}=0$ (free-slip) at the bottom ($z=0$), and $\dv*{\overline{u}}{z}=0$ at the surface ($z=1$).

As a comparison, Fig.~\ref{fig:single_stem}$(a)$  presents the solution of  \citet{wong2020shear} for neutrally-buoyant blades with flexural rigidity $EI$ and a free-slip bottom boundary condition, where the parameter controlling the blade deformability is the Cauchy number $C_Y=\rho b C_D H^3 U^2/EI$.
In our model, $EI\to0$, $C_Y\to\infty$, buoyancy is the dominant agent that resists the fluid drag on the blade, and therefore $\beta$ quantifies to what extent the blade can deform. For these steady-state solutions, $\beta$ is varied while keeping $Re$, $\lambda$ and $r$ constant. Note that $\beta$ and $Re$ can be adjusted without modifying any of the other dimensionless groups by tuning the blade dimension $d$ and $\nu^*$, respectively.

\thispagestyle{empty}

%Fig.~\ref{fig:single_stem} presents the steady-state velocity profiles and grass shapes for variable grass buoyancy and fixed $Re=1000$, $r=0.5$, and $\lambda=1$.  Note that because $\overline{\pd{h}{x}}=0$, variations of $Fr$ do not impact the steady-state results (see dimensionless equation in Methods section). Fig.~\ref{fig:single_stem}$(a)$ is a reproduction of the results in~\citet{wong2020shear} for variable $C_Y$, while Figs~\ref{fig:single_stem}$(b,c)$ correspond to our model for selected $\beta$ between 0.01 to 0.10.
%Fig.~\ref{fig:single_stem}$(b,c)$ corresponds to the velocity and grass shape with a free-slip and no-slip bottom, respectively. The limiting case $\beta\to\infty$ corresponds to a fixed, vertical blade. The free-slip solution for $\beta\to\infty$ in Fig.~\ref{fig:single_stem}$(b)$ matches the solutions in \citet{singh2016linear} and \citet{wong2020shear} for $C_Y=0$, as expected.

Grass blades with bending stiffness (Fig.~\ref{fig:single_stem}$(a)$) and buoyancy (Fig.~\ref{fig:single_stem}$(b)$) result in qualitatively similar solutions, with the main difference arising from the clamped bottom boundary condition in \citet{wong2020shear} that prevents deflection, compared to the hinged bottom boundary condition we used. 
The hinged condition results in lower drag near the bottom boundary.
Another difference is the fluid velocity at the tip of the grass. For the buoyant model in Fig.~\ref{fig:single_stem}$(b)$, we observe a monotonic growth of the velocity at the tip with decreasing $\beta$, while in Fig.~\ref{fig:single_stem}$(a)$ there is a peak and then a decay for increasing $C_Y$. 
Additionally, small variations of $\beta$, as $\beta\to 0$, drastically change the steady-state solutions (note the differences for $\beta=0.01$ and 0.015 in Fig.~\ref{fig:single_stem}$(b)$, for example).

The no-slip bottom boundary condition (Fig.~\ref{fig:single_stem}$(c)$) reduces blade deflection at the root, resulting in a smaller range of grass deflection angles and ultimately less deflection at the tip.
The shape of the velocity profile near the canopy top and  the velocity shear magnitude are not sensitive to the bottom boundary condition choice.
The no-slip condition at $z=0$ is used for the time-dependent simulations as it is physically more accurate.
\thispagestyle{empty}

% \clearpage
% \section{Supplementary Information B}

% \small{
% \noindent \emph{Seagrass deformation affects fluid instability and tracer exchange in canopy flow (Vieira, Allshouse \& Mahadevan, 2022)}
% }

\clearpage
\section{Vortex merger events and tracer exchange}
\label{subsec:vortex_merges}
\thispagestyle{empty}

\begin{figure}[h!]
    \centering
    \includegraphics[width=\textwidth]{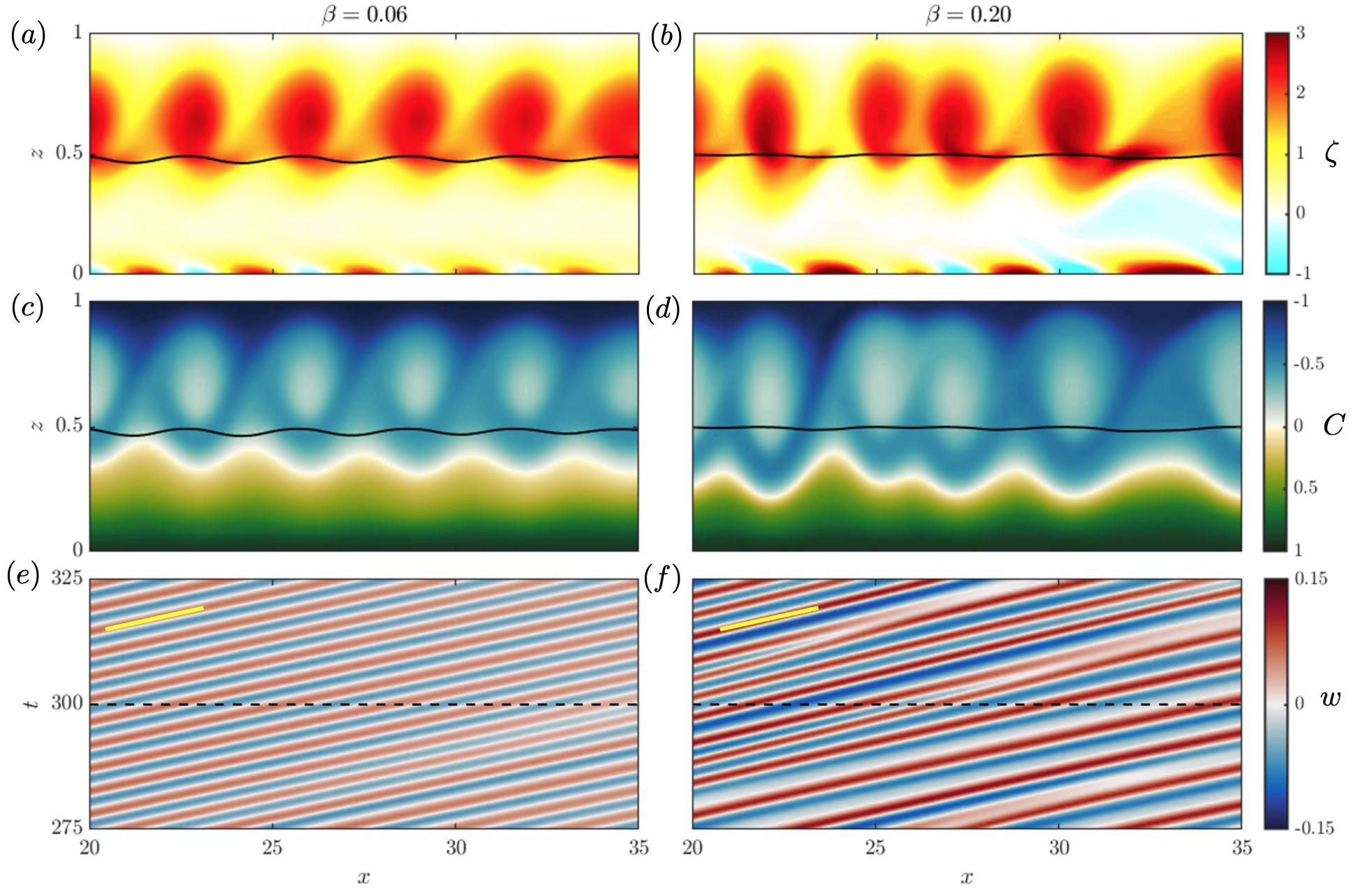}
    \caption{Vortex interaction for variable $\beta$ and fixed $Re=1500$. Instantaneous $(a,b)$ vorticity $\zeta(x,z,t=300)$ and $(c,d)$ scalar field $C(x,z,t=300)$. $(e,f)$ Hovmöller diagrams presenting the space and time evolution of $w(x,z=\overline{h}_g,t)$ for $\beta=0.06$ ($(a,c,e)$, more deformable) and $\beta=0.20$ ($(b,d,f)$, less deformable), highlighting vortex interactions and merger events for $\beta=0.20$. The yellow slope indicates the reference $\dv*{x}{t}=0.6$, and the dashed black line marks the time $t=300$ when the fields are plotted in $(a-d)$.}
    \label{fig:vortex_pairings}
\end{figure}

At combinations of high $Re$ and high $\beta$, larger-scale time fluctuations in $\Phi(t)$ become more apparent (Fig.~\ref{fig:result_exchange_grid}). 
To understand what contributes to these additional time scales to the flow, we study the vortex structure and tracer concentration fields for $Re=1500$.
The simulations presented in Fig.~\ref{fig:vortex_pairings} only vary the buoyancy parameter, with the left and right panels corresponding to $\beta=0.06$ and $\beta=0.20$, respectively.
Figs~\ref{fig:vortex_pairings}($a,b$) present the vorticity $\zeta$, and Figs~\ref{fig:vortex_pairings}($c,d$) the tracer concentration $C$, both at time $t=300$.
These plots highlight how the vorticity field lines up accurately with the tracer field at a given time, even though the tracer concentration has been evolving for $t\in[0, 300]$.

While for $\beta=0.06$ (Figs~\ref{fig:vortex_pairings}$(a,c)$) all vortices look similar and periodic, from both vorticity and tracer perspectives, the $\beta=0.20$ results (Figs~\ref{fig:vortex_pairings}$(b,d)$) show signals of vortex interaction, with an imminent vortex merger~\citep{ho1984perturbed} that has started at $x\approx 26$ (see Supplementary information for a video of the time evolution of these merger events).

Figs~\ref{fig:vortex_pairings}$(e,f)$ present Hovmöller diagrams of $w(x,z=\overline{h}_g,t)$ for both cases, similar to the ones in Figs~\ref{fig:result_onsetVisualization}$(a,b)$.
A periodic signal with vortices propagating at constant speed and amplitude is observed for the more deformable blade case~(Fig.~\ref{fig:vortex_pairings}$(e)$), while more nonlinear interactions and vortex merger events are observed for the less deformable case (Fig.~\ref{fig:vortex_pairings}$(f)$).
Both cases have a similar dominant vortex speed that is again close to the $0.6$ obtained for $\beta=0.10$ in Fig.~\ref{fig:result_onsetVisualization}$(b)$.
However, the $w$ amplitudes for $\beta=0.20$ are stronger, more variable, and the periodicity of the signal is less apparent.
This variability may be due to bigger vortices for $\beta=0.20$ growing enough to interact with neighboring vortices, causing merger events and other nonlinear phenomena that reduce periodicity in the flow. No preferred frequency for vortex merger events was observed.
\thispagestyle{empty}

\clearpage
% \section{Supplementary Information C (captions embedded in videos)}
\section{Video description}
\label{subsec:video_description}
\thispagestyle{empty}

% \vspace{1em}

The four supplementary videos, for which captions are available below, correspond to the time evolution of the results presented in Figs \ref{fig:result_onsetVisualization_field}, \ref{fig:result_betavariation}, \ref{fig:result_exchange_grid}, and \ref{fig:vortex_pairings}.

\begin{itemize}
    \item[$\star$] \texttt{Video 1}\\
    Instability onset. (top) Vorticity field $\zeta$ for the full domain. (bottom) From left to right, steady-state horizontal velocity $\overline{u}$, horizontal velocity perturbation $u'$, vertical velocity perturbation $w$, and grass blade positions $(x_g,z_g)$, for the designated region of the domain (dashed rectangle). Black lines represent the instantaneous seagrass height $h_g(x,t)$. Video corresponds to Fig.~\ref{fig:result_onsetVisualization_field}$(a,b)$ in the manuscript.
    \\
    \item[$\star$] \texttt{Video 2}\\
    Tracer transport for $\beta=0.14$. (top-left) tracer concentration $C$, (bottom-left) vertical tracer flux $\phi$, and (right) tracer exchange $\Phi$. Video corresponds to Figs~\ref{fig:result_betavariation}$(b,d)$ and \ref{fig:result_exchange_grid}$(a)$ in the manuscript.
    \\
    \item[$\star$] \texttt{Video 3}\\
    Tracer transport for (left) $\beta=0.06$ and (right) $\beta=0.14$. (top) tracer concentration $C$, (middle) vertical tracer flux $\phi$, and (bottom) tracer exchange $\Phi$. Video corresponds to Figs~\ref{fig:result_betavariation} and \ref{fig:result_exchange_grid}$(a)$ in the manuscript.
    \\
    \item[$\star$] \texttt{Video 4}\\
    Vortex merger events for $Re=1500$ and $\beta=0.20$. (top) Vorticity field $\zeta$ and (bottom) tracer concentration $C$. Video corresponds to Fig.~\ref{fig:vortex_pairings}$(b,d)$ in the manuscript.
    
\end{itemize}

\end{document}